\documentclass[10pt,a4paper,twocolumn]{article}

\usepackage{cite}
\usepackage{amsmath,amssymb,amsfonts}
\usepackage{algorithmic}
\usepackage[linesnumbered, ruled,vlined]{algorithm2e}

\usepackage{graphicx}
\usepackage{textcomp}

\usepackage{booktabs} 

\usepackage{comment}
\usepackage{balance} 
\usepackage{booktabs}

\usepackage{blindtext}
\usepackage{wrapfig}
\usepackage{amsmath}
\usepackage[dvipsnames,svgnames,x11names]{xcolor}
\usepackage{soul}  
\definecolor{darkviolet}{rgb}{0.58, 0.0, 0.83}
\definecolor{chartreuse(web)}{rgb}{0.5, 1.0, 0.0}
\definecolor{darkpastelgreen}{rgb}{0.01, 0.75, 0.24}
\definecolor{bittersweet}{rgb}{1.0, 0.44, 0.37}

\usepackage{array}
\usepackage{ragged2e}
\newcolumntype{P}[1]{>{\RaggedRight\hspace{0pt}}p{#1}}
\newcolumntype{Z}[1]{>{\centering\arraybackslash\hspace{0pt}}p{#1}}

\usepackage{pbox}

\usepackage{enumitem}

\usepackage{lineno}
\usepackage{authblk}
\usepackage{graphicx}
\usepackage[explicit]{titlesec}
\usepackage[labelfont=bf,labelsep=endash,font=footnotesize]{caption}
\usepackage{tabu}
\usepackage{xcolor}
\usepackage[hidelinks, bookmarks=false]{hyperref}  %remove bookmarks
\usepackage[hang]{footmisc}
\usepackage[normalem]{ulem}
\usepackage[top=2.5cm, bottom=2.8cm, left=1.5cm, right=1.5cm]{geometry}
\usepackage{abstract}
\usepackage{mathtools}
\usepackage{balance}

\usepackage{xurl}

\usepackage{caladea}

\makeatletter
\renewcommand\AB@affilsepx{, \protect\Affilfont}
\makeatother

\providecommand{\keywords}[1]{\textbf{Keywords}\ \ \textendash\ \   #1}

\titleformat{\section}{\large\bfseries}{\thesection.}{1em}{\MakeUppercase{#1}}
\titlespacing*{\section}{0pt}{12pt}{6pt}

\titleformat{\subsection}{\large}{\thesubsection}{1em}{#1}
\titlespacing*{\subsection}{0pt}{12pt}{6pt}

\titleformat{\subsubsection}{\large\itshape}{\thesubsubsection}{1em}{#1}
\titlespacing*{\subsubsection}{0pt}{12pt}{6pt}

\newcommand{\ITUurl}[1]{\textcolor{blue}{\urlstyle{same}\url{#1}}}

\newcommand{\ITUnote}[1]{\begin{small} \ITUpar NOTE: #1 \end{small}}

\newcommand{\ITUpar}{\vspace{8pt}\par}

\renewenvironment{abstract}
               {\list{}{
               \setlength{\rightmargin}{0mm}
               \setlength{\leftmargin}{0mm}
               \vspace{-0.25in}
                \item[\textit{\textbf{\hspace{22pt}Abstract  }}  \textendash]\relax}}
               {\endlist}

\def\starttable{\vspace{6pt}\begin{table}[ht]\center}
\def\startfigure{\vspace{6pt}\begin{figure}[ht]\center}

\makeatletter
\def\tagform@#1{\maketag@@@{\ignorespaces#1\unskip\@@italiccorr}}
\makeatother

\title{\large{\textbf{\uppercase{3-of-3 Multisignature Approach for Enabling Lightning Network Micro-payments on IoT Devices*}}}}

\def\correspondingauthor{\ITUnote{Corresponding author: Ahmet Kurt (akurt005@fiu.edu) 

*A preliminary version of this paper was accepted as a poster paper to 2021 IEEE International Conference on Blockchain and Cryptocurrency (IEEE ICBC 2021).}}

\author[1]{\normalsize{Ahmet Kurt}}
\author[1]{\normalsize{Suat Mercan}}
\author[2]{\normalsize{Enes Erdin}}
\author[1]{\normalsize{Kemal Akkaya}}

\affil[1]{\normalsize{Department of Electrical and Computer Engineering, Florida International University, Miami, FL 33174, United States}}
\affil[2]{\normalsize{Department of Computer Science, University of Central Arkansas, Conway, AR 72035, United States}\correspondingauthor{}}
\date{\vspace{-12pt}\endgraf\rule{\textwidth}{1pt}}

\begin{document}

\pagenumbering{gobble}
% \linenumbers

\twocolumn[

\begin{@twocolumnfalse}
\maketitle

\begin{abstract}
\textit{
%%%%MOTIVATION%%%%
Bitcoin's success as a cryptocurrency enabled it to penetrate into many daily life transactions. Its problems regarding the transaction fees and long validation times are addressed through an innovative concept called the Lightning Network (LN) which works on top of Bitcoin by leveraging off-chain transactions. This made Bitcoin an attractive micro-payment solution that can also be used within certain IoT applications (e.g., toll payments) since it eliminates the need for traditional centralized payment systems.
%%%CHALLENGE%%%%%
Nevertheless, it is not possible to run LN and Bitcoin on resource-constrained IoT devices due to their storage, memory, and processing requirements. 
%%%SOLUTION%%%
Therefore, in this paper, we propose an efficient and secure protocol that enables an IoT device to use LN's functions through a gateway LN node even if it is not trusted. The idea is to involve the IoT device only in signing operations, which is possible by replacing LN's original 2-of-2 multisignature channels with 3-of-3 multisignature channels. Once the gateway is delegated to open a channel for the IoT device in a secure manner, our protocol enforces the gateway to request the IoT device's cryptographic signature for all further operations on the channel such as sending payments or closing the channel. LN's Bitcoin transactions are revised to incorporate the 3-of-3 multisignature channels. In addition, we propose other changes to protect the IoT device's funds from getting stolen in possible revoked state broadcast attempts.
%%%EVALUATION AND RESULTS%%%%
We evaluated the proposed protocol using a Raspberry Pi considering a toll payment scenario. Our results show that timely payments can be sent and the computational and communication delays associated with the protocol are negligible.
}
\end{abstract}

\ITUpar

\ITUpar
\ITUpar

\end{@twocolumnfalse}
]

\section{Introduction}
The Internet of Things (IoT) has been adopted in various domains at a great pace in the last decade as it brings numerous opportunities and convenience \cite{li2015internet}. In such applications, typically resource-constrained IoT devices supply data from their sensors to remote servers through a wireless connection. With their increased capabilities, we have been witnessing applications where an IoT device may need to do financial transactions. For instance, IoT devices may be used in commercial applications such as toll systems, where an on-board unit acting as an IoT device on a vehicle may need to do automatic payments as the vehicle passes through a toll gate \cite{pavsalic2016vehicle}. Similarly, there are other cases such as automated vehicle charging, parking payment, sensor data selling, etc. where \textit{micro-payments} need to be made \cite{mercan2021cryptocurrency, kurt2020lnbot}. 

In these applications, the common feature is device-to-device (D2D) communication which may not involve any human intervention. Therefore, transactions should be automated. While these automated payments may be linked to credit card accounts of device owners, this is not only inconvenient but also requires the involvement of third parties that will bring additional management overhead. In this context, cryptocurrencies have great potential to provide a smooth payment automation. Thus, a successful merge of IoT and cryptocurrency technologies such as Bitcoin \cite{nakamoto2019bitcoin} and Ethereum \cite{wood2014ethereum} looks promising.

However, despite their popularity, mainstream cryptocurrencies such as Bitcoin and Ethereum suffer from scalability issues in terms of transaction confirmation times and throughput \cite{zhou2020solutions}. This increases the transaction fee and makes their adoption infeasible for micro-payments. The Payment Channel Network (PCN) idea has emerged as a second layer solution to address this problem by utilizing off-chain transactions \cite{decker2015fast}. For instance, Lightning Network (LN) \cite{poon2016bitcoin} is the PCN solution designed for Bitcoin which exceeded 20,000 nodes in three years. While LN is a successful solution, the current LN protocol cannot be run on most IoT devices because of the computation, communication, and storage requirements \cite{lniotproblem}. As well known, IoT devices are mostly resource-constrained and most of them are not capable of running the LN protocol where a full Bitcoin node (e.g., as of today 349 GB of storage area is required) has to be running alongside an LN node. Additionally, a robust Internet connection and relatively high computation power are required to receive and verify new blocks for the Bitcoin node. Even if we can empower some IoT devices with the needed resources, these IoT devices still need to be always online to receive synchronization messages from both Bitcoin and LN, which is not realistic for IoT either.

Therefore, there is a need for a lightweight solution that will enable resource-constrained IoT devices to utilize LN for micro-payments. To this end, in this paper, we propose an efficient and secure protocol where an IoT device can connect to an \textit{untrusted LN gateway} that already hosts the full LN and Bitcoin nodes and can: 1) open/close LN channels and 2) send payments on behalf of the IoT device when requested. Our approach is similar to a delegation approach which comes with almost negligible computation and communication overheads for the IoT devices. We are proposing to incentivize the LN gateway to participate in this payment service by letting it charge IoT devices for each payment it processes.

In our proposed protocol, we introduce the concept of \textit{3-of-3 multisignature LN channels}, which involves signatures of all three channel parties (i.e., the IoT device, the LN gateway, and a bridge LN node to which the LN gateway opens a channel for the IoT device) to conduct any operation on the channel as opposed to using the LN's original 2-of-2 multisignature channels. This modification to the channels is possible by changing the LN's Bitcoin scripts which play a critical role in our protocol as it prevents the LN gateway from spending the IoT device's funds. More specifically, the LN gateway cannot spend the IoT device's funds in the channel without getting the IoT device's cryptographic signature which consequently means that the IoT device's funds are secure at all times. Since LN's original protocol is modified, this also necessitates revisiting the \textit{revoked state broadcast} issue of LN. We offer revisions to protect the IoT device's funds in revoked state broadcast attempts when 3-of-3 multisignature LN channels are used.

To assess the effectiveness and overhead of the proposed protocol, we implemented it within a setup where a Raspberry Pi sends LN payments to a real LN node through a wireless connection. We considered two real-life cases in the experiments: a vehicle at a certain speed making a toll payment through a wireless connection and a customer paying for a coffee at a coffee shop using his/her smartwatch. We demonstrated that the proposed protocol enables the realization of timely payments with negligible computational and communication delays. We separately provide a security analysis of the proposed protocol.

The structure of the rest of the paper is as follows. In Section \ref{sec:RelatedWork}, we provide the related work. Section \ref{sec:background} describes the LN, its underlying mechanisms and its protocol specifications. System and threat model are explained in Section \ref{sec:systemmodel}. We explain the proposed protocol in detail in Section \ref{sec:protocol}. Security analysis against the assumed threats is provided in Section \ref{sec:threatanalysis}. We present our evaluation results for the proposed protocol in Section \ref{sec:evaluation}. Finally, the paper is concluded in Section \ref{sec:conclusion}.

\section{Related Work}
\label{sec:RelatedWork}

The closest work to ours is from Hannon and Jin \cite{hannon2019bitcoin}. They propose to employ LN-like payment channels to give IoT devices the ability to perform off-chain transactions. Since an IoT device does not have access to the blockchain, they use a pool of two different third parties which are called the IoT payment gateway and watchdog to aid the IoT device in the process. Using game theory, they show that the protocol reaches an equilibrium given that the players follow the protocol. However, this approach has a major issue: They assume that the IoT device can actually open LN-like payment channels to the IoT gateway. While it is not clear how it can be done, the authors also do not have a proof of concept implementation of the approach which is another puzzling aspect of the work.

Robert et al. \cite{robert2020enhanced} proposed IoTBnB, a digital IoT marketplace, to let data trading between the data owners and the users. In their scheme, after the user selects which item to buy from the marketplace, it is redirected to an LN module that performs the payment. This LN module is hosting the full Bitcoin and LN nodes. However, the authors are focused on integrating LN into an existing IoT ecosystem rather than enabling individual IoT devices to use LN that are not part of such an ecosystem. Additionally, in their system, the IoT device's funds are held by wallets belonging to the ecosystem which raises security and privacy concerns. In our work, we cover a broader aspect of IoT applications and IoT's funds are not held by third parties.

A different work focusing on Ethereum micro-payments rather than Bitcoin was proposed by Pouraghily and Wolf \cite{pouraghily2019lightweight}. They employ a Ticket-Based Verification Protocol (TBVP) for a similar purpose, enabling IoT devices to perform financial transactions in an IoT ecosystem. By using two entities called contract manager and transaction verifier, attempts are made to reduce the performance requirements on the IoT devices. There are some issues with this approach: 1) It is mentioned that the IoT funds are deposited into an account jointly opened with a partner device. Since the details are not provided, it raises security and privacy concerns. 2) The scheme utilizes Ethereum smart contracts and TBVP was compared with $\mu$Raiden \cite{raiden} which is an Ethereum payment channel framework targeting low-end devices. However, this comparison might not be reliable as $\mu$Raiden development was dormant for more than 2 years. In our work, we targeted Bitcoin and LN which are actively being developed and currently dominating the market.

A recent work by Profentzas et al. \cite{profentzas2020tinyevm} proposes TinyEVM, an Ethereum based off-chain smart contract system to enable IoT devices to perform micro-payments. The authors tried to tackle the problem by modifying Ethereum virtual machine and running it on the IoT device. In our approach, we only require the IoT device to generate signatures, which is not a resource-intensive operation. Another work, \cite{li2020data}, focuses on data transactions for IoT using payment channels. A slightly different work by authors of \cite{tapas2020p4uiot} explores using LN for delivering patch updates to the IoT devices. They employ LN in the process of claiming rewards upon successful delivery of the patch updates.

There have been commercialized implementation efforts to create lighter versions of LN for low-resource devices. Neutrino \cite{neutrino} is one of them which is a Bitcoin light client specifically designed for LN. The idea is to use the block headers only as opposed to using the whole blockchain. Breez \cite{breez} is another example which a mobile client based on lnd \cite{lnd} and Neutrino. While a portion of the IoT devices might be able to run these software, they still need to be online to synchronize block headers. Thus, we propose a solution that does not require staying online all the time or synchronizing any messages after coming online.

This work is an extension to our poster paper \cite{kurt2021enabling} with a lot of new content. 1) The poster version does not have the full details of the proposed protocol. In this paper, we explain the protocol in detail and show the modifications to LN's existing specifications. 2) The poster version does not have related work and background sections. In this paper, we give comprehensive background information on LN and provide the related work. 3) In this journal version, we present the threat model and the security analysis that did not exist in the poster version. 4) The evaluation section in the poster version only has the WiFi experiments. In this paper, we present Bluetooth experiments in addition to the WiFi. Additionally, we present another use-case scenario in the experiments and provide a cost analysis.

\section{Background}
\label{sec:background}
In this section, we provide comprehensive background information on LN, its components and specifications to help understand our approach.

\subsection{Lightning network}

LN was introduced by Poon and Dryja in 2015 in a technical paper \cite{poon2016bitcoin}. After 2 years of its introduction, it was implemented by Lightning Labs and started being used on Bitcoin mainnet \cite{lnlaunch}. The intuition behind developing LN which is a layer-2 payment channel network is to solve Bitcoin's \textit{scalability} problem. Similar to Bitcoin, LN is a peer-to-peer distributed network but it is not standalone, rather it is operating on top of Bitcoin. The idea of creating networks on top of blockchains is not new \cite{decker2015fast}. For Bitcoin, it is possible to create and deploy such networks like LN by utilizing its \textit{smart contract} feature \cite{smartcontract}. In this way, secure \textit{payment channels} can be established which can be used by users for instant and almost free Bitcoin transactions. When enough of these payment channels are opened by the users, they form a network of payment channels. Then, this network can be utilized by new users to route their payments to specific destinations. Such payments where the existing channels on LN are utilized are called multi-hop payments. This feature of LN eases the onboarding process for new users to start using the network for sending/receiving payments.

\begin{figure}[h]
    \centering
    \includegraphics[width=\linewidth]{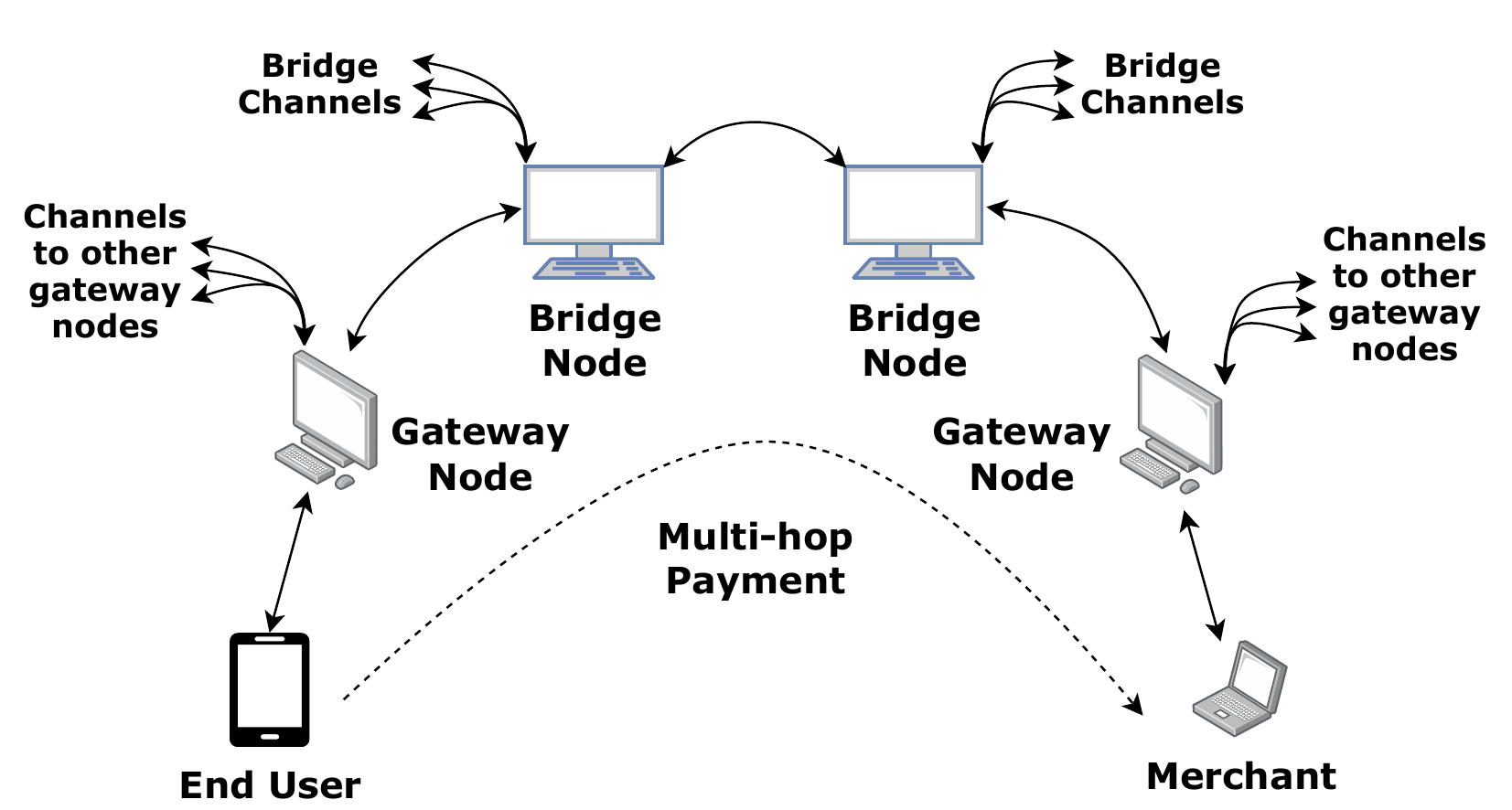}
    \caption{An illustration of the topology of LN adapted from \cite{lntopology}. Payments can be routed over existing channels.}
    \label{fig:lntopology}
\end{figure}

While LN is mostly serving end users and merchants, there are other entities within it that serve different purposes. For example, an end user trying to connect to the LN using a mobile device will be connecting to a \textit{gateway node}. Such gateway nodes directly serve the end users and earn fees by routing the payments. After connecting to the LN, the end user's payments have to be conveniently routed to other participants of the network. \textit{Bridge nodes} serve this purpose. They enable the connectivity between the existing gateway nodes and also earn routing fees. Even though they have different names, bridge nodes and gateway nodes are essentially all regular LN nodes distinguished by their roles in the network. Fig. \ref{fig:lntopology} summarizes the virtual topology of the LN. The network is highly scalable and can support millions of transactions per second. After its creation, LN grew exponentially reaching 21,310 nodes maintaining 48,915 channels at the time of writing this paper \cite{1ml}.

\begin{figure*}[h]
  \centering
  \vspace{-3mm}
  \includegraphics[width=0.6\linewidth]{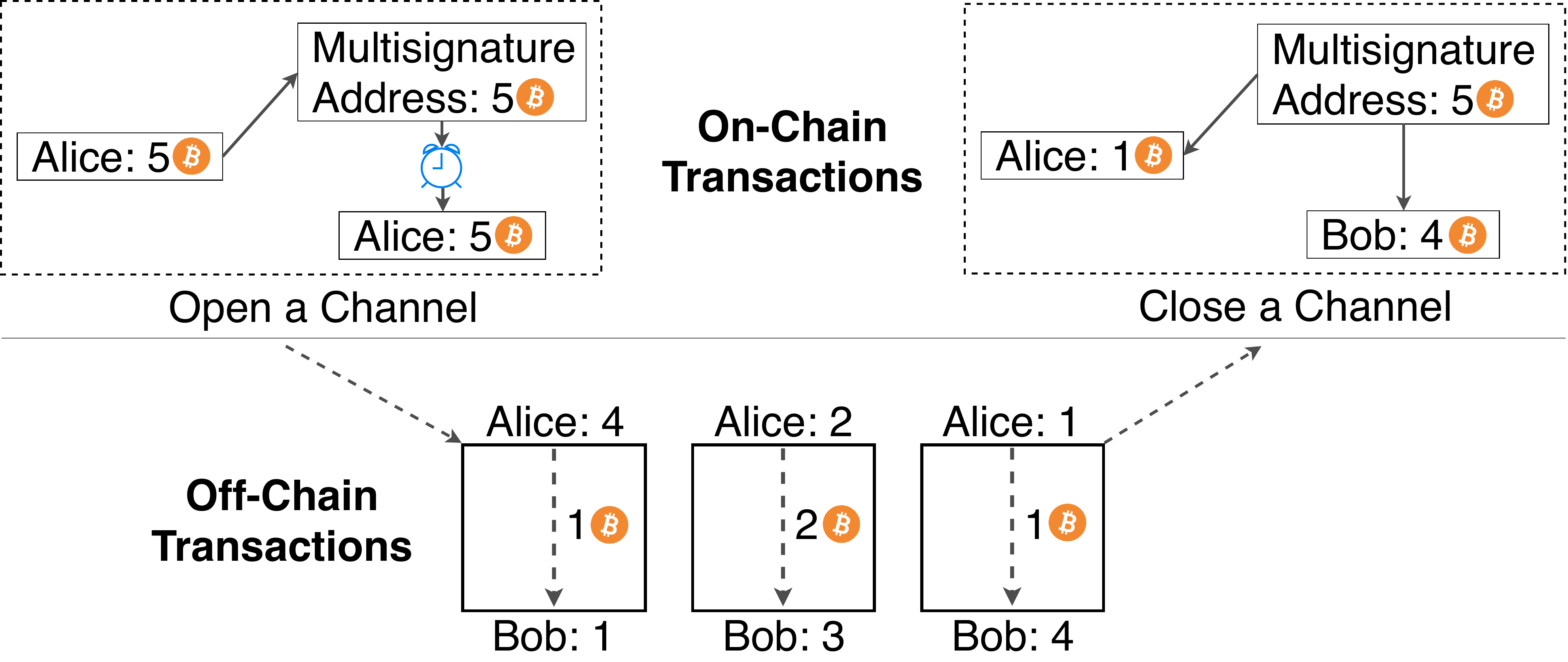}
  \caption{A depiction of the on-chain channel opening and closing transactions and off-chain payments in LN.}
  \vspace{-2mm}
  \label{fig:offchain}
\end{figure*}

\subsection{Underlying LN mechanisms}
\label{sec:LNmechanisms}

In this section, we will briefly touch upon the key concepts of LN which are crucial to understanding the protocol descriptions. To explain these concepts, we use an example case where Alice opens an LN channel to Bob with the purpose of sending him LN payments.

\vspace{1mm}
\noindent \textit{Funding transaction}: When Alice wants to open a channel to Bob, she needs to construct a proper funding transaction first. This on-chain transaction determines the channel capacity which is the amount of funds that will be committed to the channel. Once Alice creates the transaction, she sends the outpoint\footnote{Transaction outpoint is the combination of the transaction output and the output index.} to Bob. Receiving the outpoint, Bob can send a signature to Alice which will enable her to broadcast the funding transaction to the Bitcoin network. In Fig. \ref{fig:offchain}, Alice opens a channel to Bob with a 5 Bitcoin capacity. Once funds are committed to the channel, she can send off-chain payments to Bob up to a total of 5 Bitcoins.

\begin{figure}[h]
    \centering
    \includegraphics[width=\linewidth]{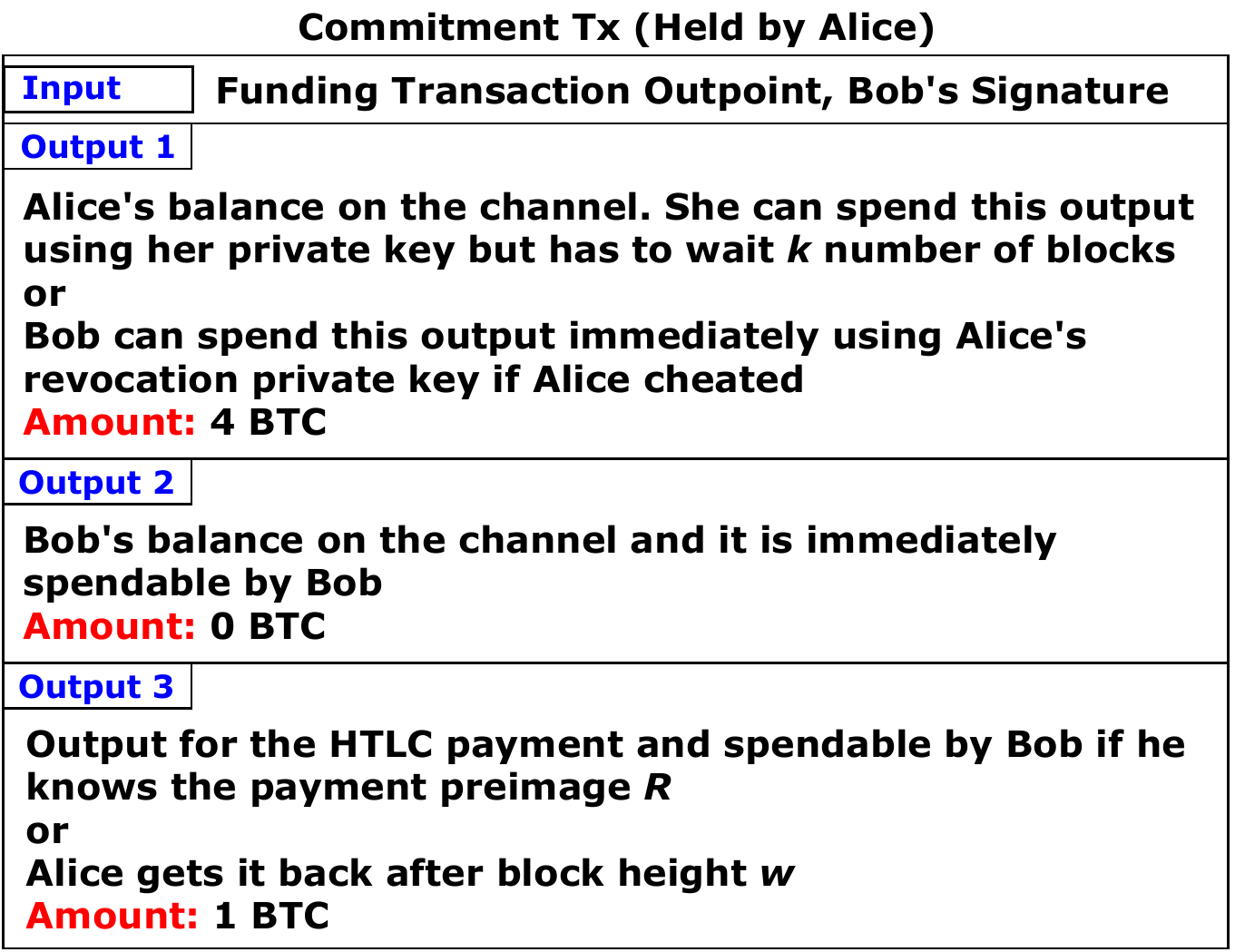}
    \caption{An illustration of the commitment transaction stored at Alice.}
    \label{fig:commitment1}
\end{figure}

\vspace{1mm}
\noindent \textit{Commitment Transaction}: When Alice starts sending Bob off-chain payments, her and Bob's balances on the channel will change. In Fig. \ref{fig:offchain}, Alice sends 3 different payments to Bob with amounts 1 Bitcoin, 2 Bitcoins, 1 Bitcoin respectively. Since these transactions are not on-chain (i.e. not mined by miners thus not included in the blocks), there has to be a different mechanism to keep track of each parties' balances in the channel. This is done by the \textit{commitment transactions}. A commitment transaction is a type of Bitcoin transaction specifically designed for LN. A payment channel consists of states, changing with each payment. In each state, parties have different balances which are recorded onto their commitment transactions. We illustrated Alice's commitment transaction in Fig. \ref{fig:commitment1} after she initiates a payment of 1 Bitcoin to Bob. The inputs to her commitment transaction are: 1) funding transaction outpoint and a signature from Bob. She receives Bob's signature for each new state which enables her to broadcast her commitment transaction if required. As for the outputs, there are 3 of them as shown in Fig. \ref{fig:commitment1}. Output 1 is for Alice's balance on the channel. If she broadcasts her commitment transaction for any reason, she can claim her funds from this output after waiting \textit{k} number of blocks. Output 2 is for Bob's balance on the channel which Bob can spend immediately. Finally, Output 3 is for the 1 Bitcoin payment Alice sent to Bob. We will explain it next.

\vspace{1mm}
\noindent \textit{HTLCs}: Hash Time Locked Contracts (HTLCs) are a core part of LN and they enable sending conditional payments. As the name suggests, an HTLC payment is hash and time locked. This means that the recipient of the payment has to redeem it within a certain period of time by revealing a secret otherwise the payment is returned to the sender. When Alice initiates a payment of 1 Bitcoin to Bob, she creates a new commitment transaction and adds an additional output to it called the \textit{HTLC output} (Output 3 in Fig. \ref{fig:commitment1}). The steps taken in this process are as follows: Alice asks Bob to generate a secret called \textit{preimage}. Bob takes a hash of the preimage and sends the hash to Alice. Alice creates an HTLC using this hash and sends the HTLC to Bob. Receiving the HTLC, Bob reveals the preimage to Alice to prove that he was the intended recipient of the payment. This finalizes the payment. The important detail here is that, if Bob does not reveal the preimage on time, 1 Bitcoin will be returned to Alice.

\vspace{1mm}
\noindent \textit{Revoked State Broadcast}: We briefly mentioned in the commitment transaction part that, the first output of Alice's commitment transaction is conditional. If Alice broadcasts her commitment transaction, she can redeem her funds from Output 1 only after waiting \textit{k} number of blocks, in other words, her funds are timelocked. This mechanism of LN is to prevent possible cheating attempts that might arise from broadcasting old commitment transactions. If Alice broadcasts a revoked (old) state, Bob can sweep Alice's funds in the channel (Output 1) by using the \textit{revocation private key} of the respective channel state while Alice is waiting to redeem the funds for the duration of the timelock. Publishing an old state is tempting for Alice since she initially had 5 Bitcoins in the channel which is more than what she has now (i.e. 4 Bitcoins). However, this cheating attempt will result in Alice losing all her funds in the channel. Therefore, channel parties are disincentivized from cheating with this timelock mechanism in the commitment transactions.

\subsection{Basis of lightning technology}
Basis of Lightning Technology (BOLT) documents specify the LN's layer-2 protocol completely. Different implementations of LN follow these specifications to be compatible with each other. We propose changes to BOLT \#2 and BOLT \#3 to implement our protocol. These two specifications are explained below.

\subsubsection{BOLT \#2: Peer protocol for channel management}
\label{sec:bolt2}

This protocol \cite{bolt2} explains how LN nodes handle channel related operations thus called the peer protocol for channel management. An LN channel has three phases which are \textit{channel establishment}, \textit{normal operation}, and \textit{channel closing}. We propose modifications to this protocol to incorporate the IoT device in channel operations. High-level explanations related to each channel phase are given below.

\vspace{1mm}
\noindent \textit{Channel Establishment}: To establish a channel, the funding node first sends an \textit{open\_channel} message to the fundee, who replies with an \textit{accept\_channel} message. Then, the funder creates the funding and the commitment transactions. It signs the fundee's commitment transaction and sends the signature to the fundee in a \textit{funding\_created} message. The fundee also sends its signature to the funder in a \textit{funding\_signed} message. Then, the funder broadcasts the funding transaction to open the channel which becomes usable after exchanging the \textit{funding\_locked} messages.

\vspace{1mm}
\noindent \textit{Normal Operation}: Now that the channel is opened, it can be used to send/receive payments. Both nodes can offer each other HTLCs with \textit{update\_add\_htlc} messages. The sender of the payment first updates the commitment transaction of the receiver, signs it, and sends the signature in a \textit{commitment\_signed} message. Receiving the signature, the receiver applies the changes and sends a \textit{revoke\_and\_ack} message to the sender to revoke the old state. Finally, the sender also applies the changes to its own commitment transaction which completes the payment.

\vspace{1mm}
\noindent \textit{Channel Close}: Both nodes can close the channel when they want. To initiate a mutual channel closing, one of the nodes sends a \textit{shutdown} message which is replied by a \textit{shutdown} message by the counterparty. This means no new HTLCs are going to be accepted. Then, the nodes start negotiating on a channel closing fee through \textit{closing\_signed} messages until they both agree on the same fee. Once negotiated, the mutually agreed channel closing transaction is broadcast to the Bitcoin network and the channel is closed. Nodes can also choose to close a channel by broadcasting their latest commitment transaction. However, this is not the preferred way to close a channel in LN and is only recommended to be done when the counterparty is not responding.

\subsubsection{BOLT \#3: Bitcoin transaction and script formats}
\label{sec:bolt3}

This specification \cite{bolt3} explains the format of LN's Bitcoin on-chain transactions. Following this protocol is essential as it ensures that the generated signatures are valid. The transactions that are specified in this protocol are: 1) the funding transaction, 2) the commitment transactions and 3) the HTLC transactions. Before explaining them, we first describe some terminology related to Bitcoin.

\vspace{1mm}
\textit{SegWit}: Segregated Witness (SegWit) was a Bitcoin soft-fork that became active in 2017. It fixed the transaction malleability problem. Before SegWit, it was possible to change the transaction ID (txid) of a transaction after it was created. Eliminating malleability also enabled deploying LN onto Bitcoin.

\vspace{1mm}
\textit{P2WSH}: Pay-to-Witness-Script-Hash (P2WSH) is a type of Bitcoin transaction that uses SegWit. Before SegWit upgrade, Bitcoin was using Pay-to-Script-Hash (P2SH) transactions.

\vspace{1mm}
\textit{Witness}: It is the part of a SegWit transaction that is not included when the transaction is hashed and signed. Thus, witness does not affect the txid.

\vspace{1mm}
\textit{Witness Script}: This is the script that describes the conditions to spend a P2WSH output.

\vspace{1mm}
\textit{Bitcoin Opcodes}: Operation codes (opcodes) specify the commands to be performed in computer languages. For Bitcoin, its scripting language has a number of opcodes that define various functions. For instance, \texttt{OP\_CHECKSIG} checks whether a signature is valid.

\vspace{1mm}
Now, we briefly explain LN's Bitcoin transactions below as specified by BOLT \#3.

\vspace{1mm}
\textit{Funding Transaction Output}: This is a P2WSH output with a witness script: \texttt{2 <pubkey1> <pubkey2> 2 OP\_CHECKMULTISIG}. It sends the funds to a 2-of-2 multisignature address which is generated from \texttt{pubkey1} and \texttt{pubkey2}. \texttt{pubkey} is also referred to as \texttt{funding\_pubkey}.

\vspace{1mm}
\textit{Commitment Transaction Outputs}: A commitment transaction currently can have up to 6 types of outputs. These outputs are:

\begin{enumerate}[leftmargin=*]
    \item \texttt{to\_local} Output: This P2WSH output is for the funds of the owner of the commitment transaction and it is timelocked using \texttt{OP\_CHECKSEQUENCEVERIFY}. It can also be claimed immediately by the remote node if it knows the revocation private key.
    
    \item \texttt{to\_remote} Output: This P2WPKH output is for the funds of the remote node which is immediately spendable by him/her.
    
    \item \texttt{to\_local\_anchor} and \texttt{to\_remote\_anchor} Output: This P2WSH output was recently added to prevent cases where the commitment transaction cannot be mined by miners due to insufficient fees. With this non-timelocked output which both nodes can spend, the commitment transaction's fee can be changed for faster mining.
    
    \item Offered HTLC Outputs: This is a P2WSH output for the HTLCs that are offered to the remote node. It either sends the funds to the HTLC-timeout transaction after the HTLC timeouts or to the remote node that can claim the funds using the payment preimage or the revocation key.

    \item Received HTLC Outputs: This is a P2WSH output for the HTLCs that are received from the remote node. The remote node can claim the funds after the HTLC timeouts or by using the revocation key or; the funds are sent to an HTLC-success transaction which can be claimed by the local node with the payment preimage.
\end{enumerate}

\vspace{1mm}
\textit{HTLC-Timeout and HTLC-Success Transactions}: LN makes use of these two-stage HTLCs to protect nodes from losing any funds\footnote{An explanation on why two-stage HTLCs are needed in LN can be found at \url{https://github.com/lnbook/lnbook/issues/187}.}. Both HTLC transactions are almost identical and can be spent with a valid revocation key in case of a cheating attempt.

The information provided in this section is essential to understand the modifications we propose to BOLT \#3 in Section \ref{sec:changescripts}.

\section{System \& Threat Model}
\label{sec:systemmodel}

In this section, we present our system model and state the threat assumptions. 

\subsection{System model}

\begin{figure}[h]
    \centering
    % \vspace{-2mm}
    \includegraphics[width=\linewidth]{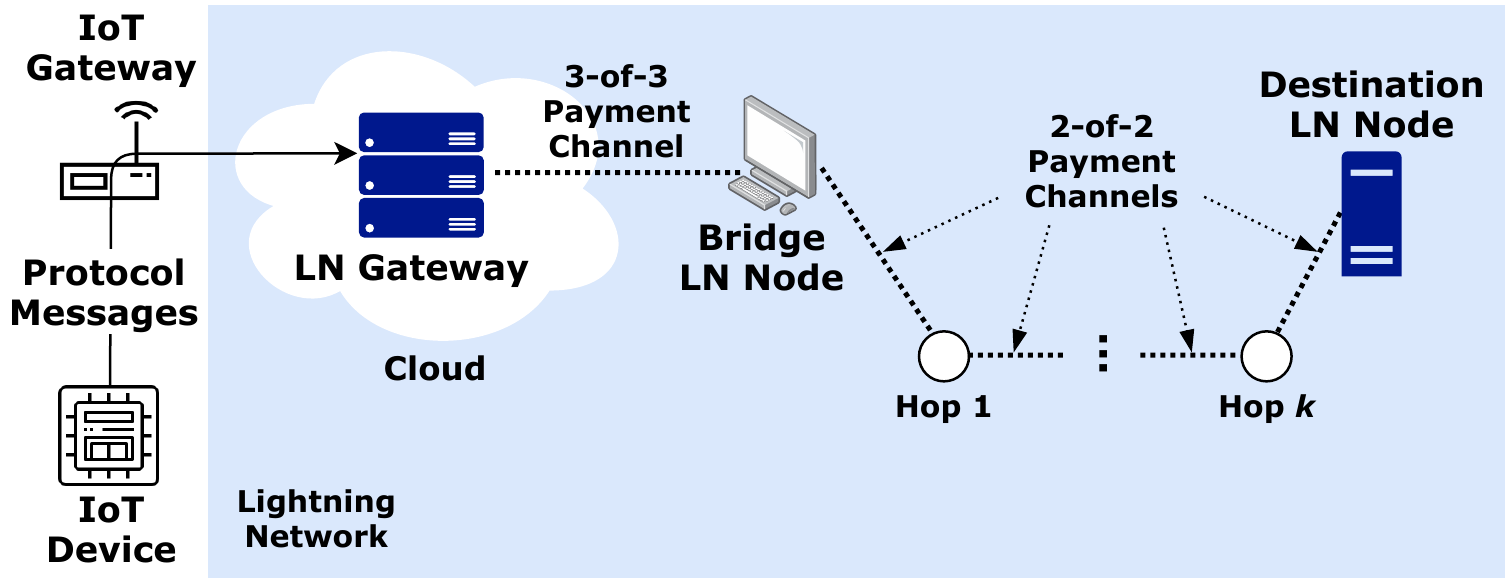}
    \vspace{-6mm}
    \caption{Illustration of our assumed system model.}
    \label{fig:system_model}
    % \vspace{-2mm}
\end{figure}

There are five entities in our system which are \textbf{IoT device}, \textbf{IoT gateway}, \textbf{LN gateway}, \textbf{bridge LN node}, and \textbf{destination LN node} as shown in Fig. \ref{fig:system_model}. The IoT device wants to send payments to the destination LN node for the goods/services. The IoT gateway acts as an access point connecting the IoT device to the Internet through a wireless communication standard such as WiFi, Bluetooth, 5G, etc. depending on the application. The IoT gateway provides connectivity when the IoT device is within its communication range. With this Internet connection, the IoT device communicates with the LN gateway which is running on the cloud. The LN gateway hosts the full Bitcoin and LN nodes to serve the IoT device and collects fees in return for its services. Upon a channel opening request from the IoT device, the LN gateway opens a channel to a bridge LN node that it selects from the existing LN nodes on the network. This bridge LN node is used to route the IoT device's payments to the destination LN node. While the LN gateway is free to choose any node, it should choose a highly connected node with many open channels to prevent routing failures. Our proposed system requires changes to the LN protocol. Therefore, the LN gateway and bridge LN node have to run the modified LN software.

We assume that both the IoT device and the LN gateway stay online during an LN operation such as sending a payment. The IoT device can be offline for the rest of the time.

\subsection{Threat model}
\label{sec:threatmodel}

We make the following security related assumptions:

\vspace{-2mm}

\begin{enumerate}[leftmargin=*]

    \item The LN gateway can post old (revoked) channel states to the blockchain. It can gather information about IoT devices and can try to steal/spend funds in the channel. However, it follows the proposed protocol specifications for communicating with the IoT device.
    
    \item The IoT device and the LN gateway have an encrypted and authenticated communication channel between them (i.e., SSL/TLS).

\end{enumerate}

We consider the following attacks to our proposed system:

\vspace{1mm}
\textbf{Threat 1: Revoked State Broadcasts:} The LN gateway and bridge LN node can broadcast revoked states to the blockchain to steal money from other parties.
    
\vspace{1mm}
\textbf{Threat 2: Spending IoT Device's Funds:} The LN gateway can spend the IoT device's funds that are committed to the channel by sending them to other LN nodes without the consent of the IoT device.

\section{Proposed Protocol Details}
\label{sec:protocol}

In this section, we explain the details of the proposed channel opening, payment sending, and channel closing protocols. As mentioned in Section \ref{sec:background}, we propose changes to LN's BOLT \#2 and BOLT \#3 and show these changes throughout the protocol description.

\subsection{Channel opening process}

As mentioned earlier, payment channels are created by the on-chain \textit{funding transaction}. In our case, the IoT device does not have direct access to Bitcoin and LN thus cannot broadcast a funding transaction by itself to open a channel. For this reason, we propose that the IoT device securely initiates the channel opening process through the LN gateway and generates signatures whenever required. This necessitates modifying the LN's existing channel opening protocol. The main modification is to require a third signature which is going to be generated by the IoT device. Regular 2-of-2 multisignature LN channels are not suitable to accompany 3 signatures; therefore, we propose to change the 2-of-2 multisignature channels to 3-of-3 multisignature channels. With this change, an LN channel can securely be opened by involving 3 parties. This is the main novelty in our approach. 

The steps of the proposed channel opening protocol are depicted in Fig. \ref{fig:openchannel}. We explain the protocol step by step below:

\begin{figure}[h]
    \centering
    \includegraphics[width=\linewidth]{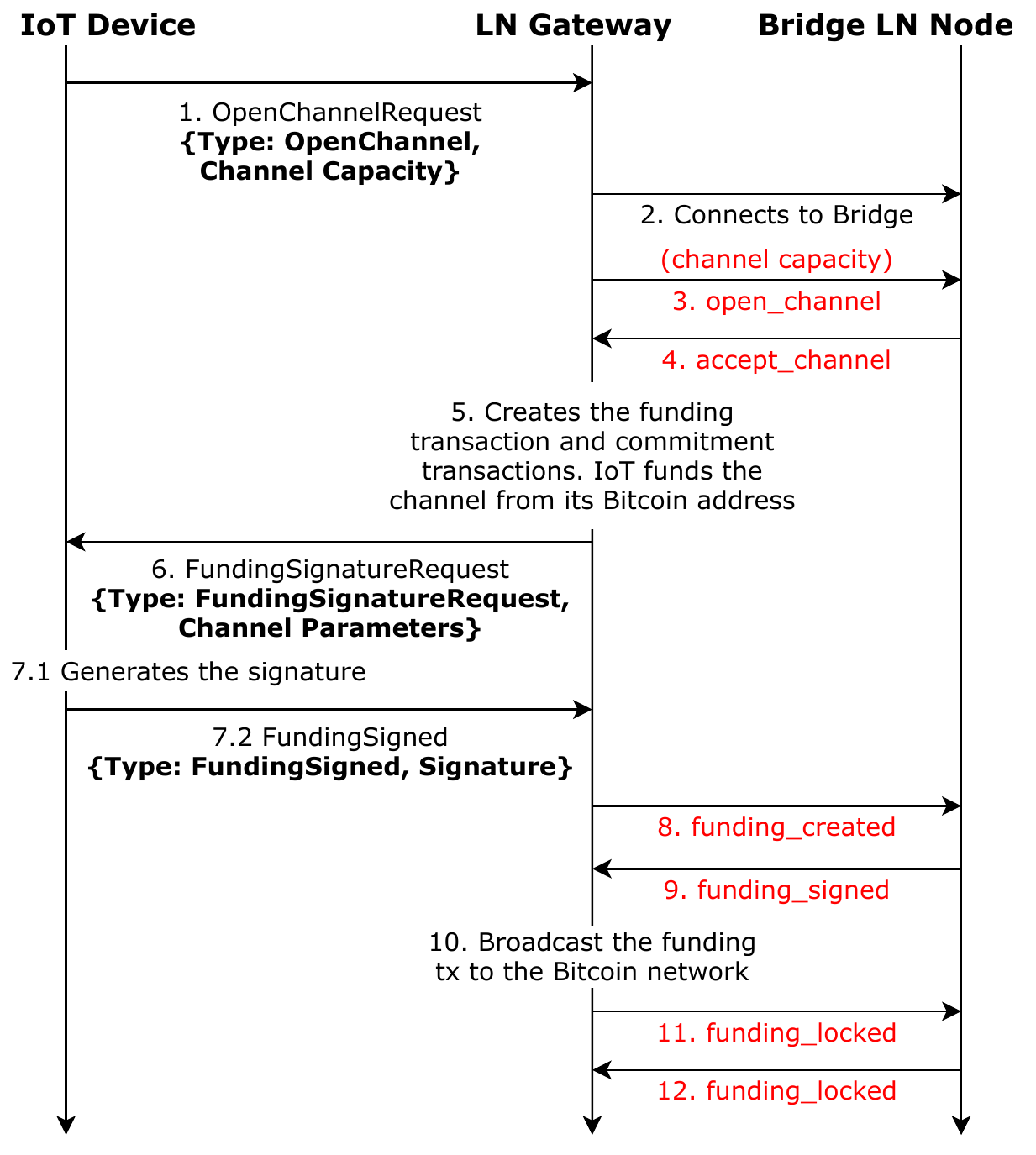}
    \caption{Protocol steps for opening a channel. Messages in red show the default messages in BOLT \#2.}
    \label{fig:openchannel}
\end{figure}

\vspace{1mm}
\textbf{IoT Channel Opening Request}: The IoT device sends an \textit{OpenChannelRequest} message to the LN gateway to request a payment channel to be opened (\#1 in Fig. \ref{fig:openchannel}). This message has the following fields: \textit{Type: OpenChannelRequest, Channel Capacity}. \textit{Channel Capacity} is specified by the IoT device and this amount of Bitcoin is taken from the IoT device's Bitcoin wallet as will be explained in the next steps.

\vspace{1mm}
\textbf{Channel Opening Initiation}: The LN gateway initiates the channel opening process upon receiving the request from the IoT device. For this, it connects to a bridge LN node which it selects from the existing LN nodes on the network. Preferably, the LN gateway chooses a node with many active channels to have good chances of getting the IoT device's payments routed to the destination LN node. Upon establishing a connection, the LN gateway sends an \textit{open\_channel} message to the bridge LN node (\#3 in Fig. \ref{fig:openchannel}). This message includes a parameter called the \textit{funding\_pubkey} which is a Bitcoin public key. Both channel parties provide their own \textit{funding\_pubkey} which are later used to construct the multisignature address of the channel. Here, we propose to add the \textit{funding\_pubkey} of the IoT device in \textit{open\_channel} message as well. Once the bridge LN node receives this message, it responds with an \textit{accept\_channel} message (\#4 in Fig. \ref{fig:openchannel}) to acknowledge the channel opening request from the LN gateway. This message includes the \textit{funding\_pubkey} of the bridge LN node.

\vspace{1mm}
\textbf{Creating the Funding and Commitment Transactions}: Exchanging these messages locks the channel parameters and the LN gateway can now start creating the funding transaction. Using the \textit{funding\_pubkeys} of all 3 parties, the LN gateway creates a 3-of-3 multisignature address which will be used to store the channel funds. Then, the LN gateway creates a Bitcoin transaction from the IoT device's Bitcoin address to the 3-of-3 multisignature address it generated. The on-chain transaction fee for this transaction is deducted from the IoT device's Bitcoin address since it requested the channel opening. As a next step, the LN gateway creates the first versions of the commitment transactions for itself and the bridge LN node. Now, the LN gateway and the bridge LN node need to exchange signatures for the newly created commitment transactions. Specifically, the bridge LN node needs IoT device's and LN gateway's signatures. Similarly, the LN gateway needs the IoT device's and bridge LN node's signatures. The process starts with the LN gateway asking for a signature from the IoT device as explained next.

\vspace{1mm}
\textbf{Getting Signature from the IoT Device}: The LN gateway now has to request a signature from the IoT device to be sent to the bridge LN node in the \textit{funding\_created} message. For this, the IoT device needs the channel parameters that were agreed on earlier between the LN gateway and the bridge LN node. Thus, the LN gateway sends a \textit{FundingSignatureRequest} message (\#6 in Fig. \ref{fig:openchannel}) to the IoT device having the following fields: \textit{Type: FundingSignatureRequest, Channel Parameters}. Having this information, the IoT device can generate the necessary signature and send it to the LN gateway in a \textit{FundingSigned} message (\#7.2 in Fig. \ref{fig:openchannel}) which has the following fields: \textit{Type: FundingSigned, Signature}.

\vspace{1mm}
\textbf{Exchanging Signatures with the Bridge LN Node}: Now that the LN gateway has the IoT device's signature, it can generate its own signature as well and send these two signatures to the bridge LN node in the \textit{funding\_created} message (\#8 in Fig. \ref{fig:openchannel}). This message also includes the funding transaction outpoint. Now, the bridge LN node is able to generate the signature for the LN gateway's commitment transaction and sends it to the LN gateway with the \textit{funding\_signed} message (\#9 in Fig. \ref{fig:openchannel}). Note that, the LN gateway does not have the IoT device's signature for its own commitment transaction yet. In case the bridge LN node becomes unresponsive at this stage, the LN gateway can ask the IoT device for its signature to close the channel unilaterally.

\vspace{1mm}
\textbf{Broadcasting the Funding Transaction}: Now is the time for the LN gateway to broadcast the funding transaction to the Bitcoin network. After broadcasting the transaction, the LN gateway and the bridge LN node wait for the transaction to reach enough depth on the blockchain (typically 3). Once the required depth is reached, they send \textit{funding\_locked} messages (\#11-12 in Fig. \ref{fig:openchannel}) to each other to lock the channel and begin using it.

\subsection{Sending a payment}

As in the case of channel opening, the introduction of the IoT device requires modifications to LN's payment sending protocol as well. We incorporate the IoT device in the process for signature generation. The details of the payment sending protocol are depicted in Fig. \ref{fig:sendpayment} and elaborated below:

\begin{figure}[h]
    \centering
    % \vspace{-2mm}
    \includegraphics[width=\linewidth]{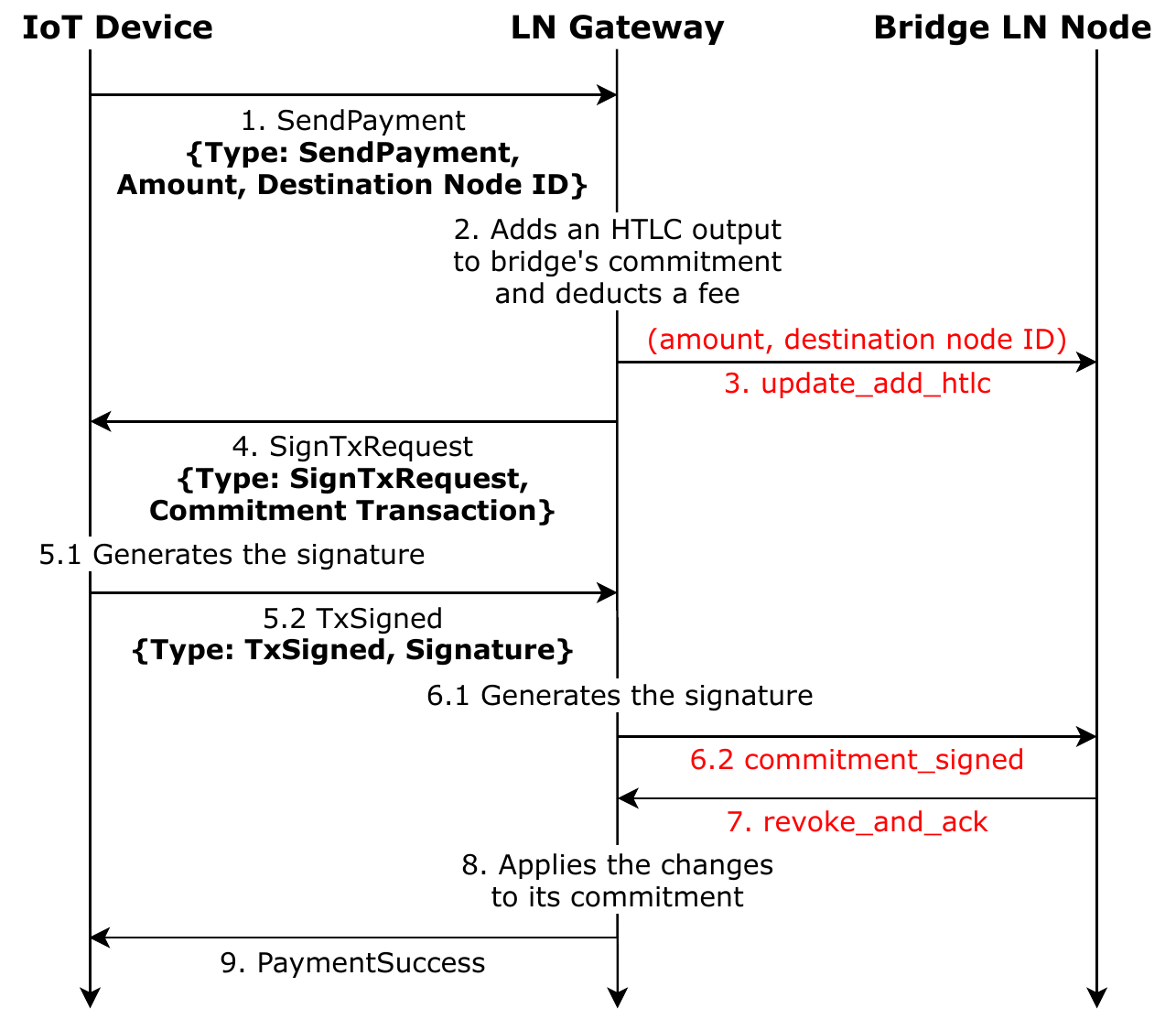}
    \vspace{-6mm}
    \caption{Protocol steps for sending a payment. Messages in red show the default messages in BOLT \#2.}
    \label{fig:sendpayment}
    % \vspace{-2mm}
\end{figure}

\vspace{1mm}
\textbf{IoT Payment Sending Request}: The IoT device sends a \textit{SendPayment} message to the LN gateway to request a payment sending (\#1 in Fig. \ref{fig:sendpayment}). This message has the following fields: \textit{Type: SendPayment, Amount, Destination Node ID}. Here, we assume that the IoT device receives the \textit{destination node ID} in some form (i.e. QR code) from the service provider (i.e. toll company).

\vspace{1mm}
\textbf{Payment Processing at the LN Gateway}: Upon receiving the request from the IoT device, the LN gateway adds an HTLC output to the bridge LN node's commitment transaction. When preparing the HTLC, the LN gateway \textit{deducts a certain amount of fee} from the real payment amount the IoT device wants to send to the destination LN node. Therefore, the remaining Bitcoin is sent with the HTLC. Then, to actually offer the HTLC, the LN gateway sends an \textit{update\_add\_htlc} message to the bridge LN node (\#3 in Fig. \ref{fig:sendpayment}). As explained in Section \ref{sec:background}, this HTLC can be redeemed with the payment preimage. \textit{Destination node ID} is embedded into the \textit{onion routing packet} in the \textit{update\_add\_htlc} message. And the \textit{amount} is also sent in this message.

\vspace{1mm}
\textbf{Getting Signature from the IoT Device}: Now, the LN gateway can apply the changes to the bridge LN node's commitment transaction and get it ready for signing. As in the case of channel opening, two signatures are needed; one from the LN gateway and one from the IoT device. Thus, we propose the LN gateway requests a signature from the IoT device for the new commitment transaction. For this purpose, the LN gateway sends a \textit{SignTxRequest} message (\#4 in Fig. \ref{fig:sendpayment}) to the IoT device which has the following fields: \textit{Type: SignTxRequest, Commitment Transaction}. The IoT device generates a signature for this commitment transaction and sends it to the LN gateway in a \textit{TxSigned} message having the following fields: \textit{Type: TxSigned, Signature} (\#5.2 in Fig. \ref{fig:sendpayment}).

\vspace{1mm}
\textbf{Exchanging Signatures with the Bridge LN Node}: Upon receiving the signature from the IoT device, the LN gateway generates its own signature as well and sends these two signatures to the bridge LN node in a \textit{commitment\_signed} message (\#6.2 in Fig. \ref{fig:sendpayment}). The bridge LN node checks the correctness of the signatures and once it verifies that the signatures are valid, it replies to the LN gateway with a \textit{revoke\_and\_ack} message (\#7 in Fig. \ref{fig:sendpayment}). This message includes the commitment secret of the old commitment transaction effectively revoking the old channel state.

\vspace{1mm}
\textbf{Payment Sending Finalization}: Now, the LN gateway can also apply the changes to its own commitment transaction. To notify the IoT device of the successful payment, the LN gateway sends a \textit{PaymentSuccess} message (\#9 in Fig. \ref{fig:sendpayment}) to the IoT device which finalizes the payment sending process.

\subsection{Channel closing process}
We briefly mentioned LN's channel closing mechanism in Section \ref{sec:bolt2}. An LN channel can be closed: 1) unilaterally when one of the channel parties broadcasts its most recent commitment transaction or 2) mutually where channel parties agree on the closing fee and create and broadcast a closing transaction. In our case, all 3 parties of the channel namely; the IoT device, the LN gateway and the bridge LN node can close the channel. We explain each case separately below:

\subsubsection{IoT device channel closure}

\begin{figure}[h]
    \centering
    % \vspace{-3mm}
    \includegraphics[width=\linewidth]{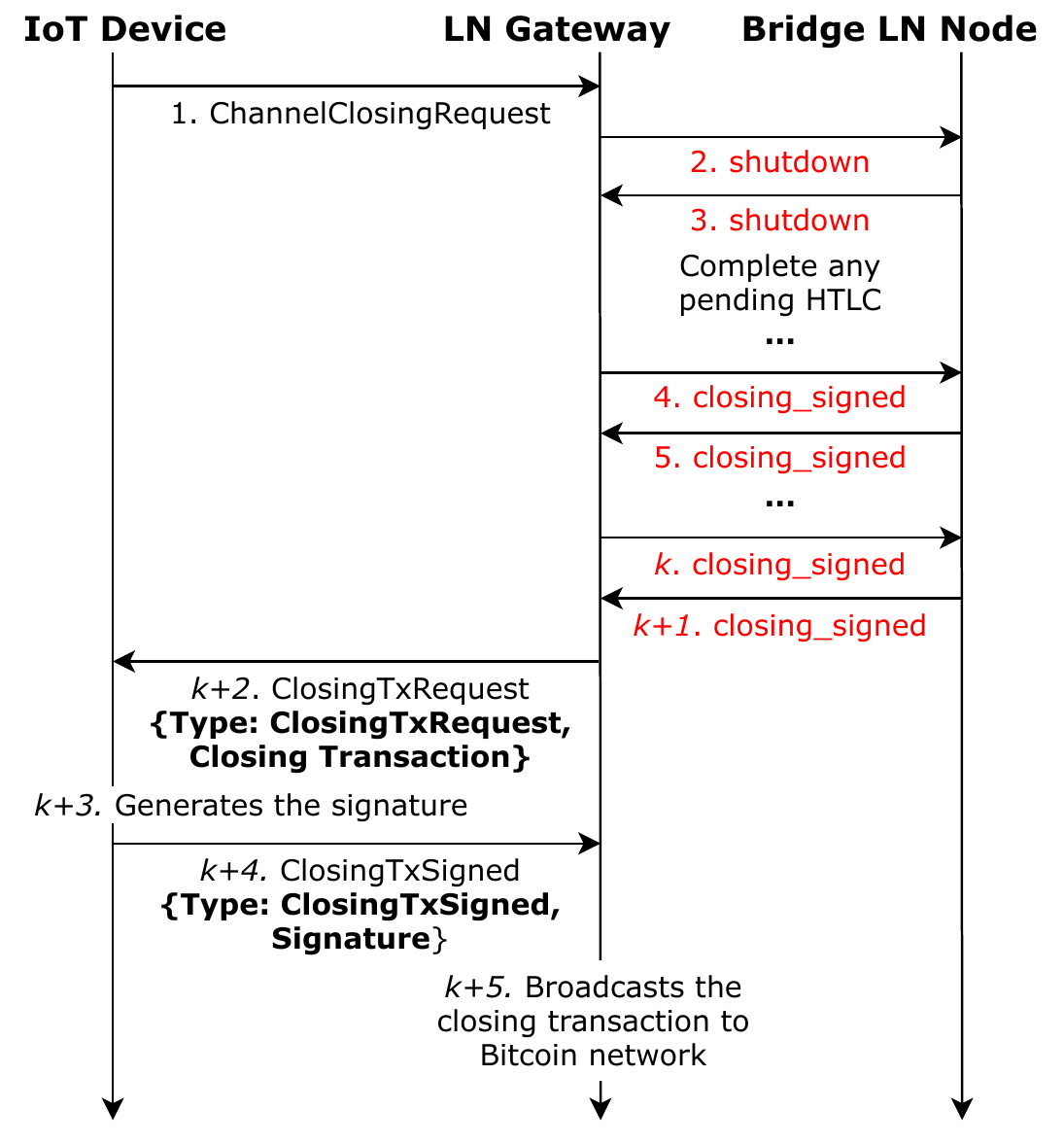}
    \vspace{-6mm}
    \caption{Protocol steps for the IoT device channel closure when the LN gateway performs a mutual close with the bridge LN node. Messages in red show the default messages in BOLT \#2.}
    \label{fig:IoTclosechannel}
    % \vspace{-2mm}
\end{figure}

To close the channel between the LN gateway and the bridge LN node which was opened due to a request from the IoT device, the IoT device follows the proposed protocol explained below, which is also shown in Fig. \ref{fig:IoTclosechannel}.

% \begin{itemize}[leftmargin=*]
    \vspace{1mm}
    \textbf{IoT Channel Closing Request}: The IoT device sends a \textit{ChannelClosingRequest} message to the LN gateway. 
    
    \vspace{1mm}
    \textbf{Mutual Close}: The LN gateway can close the channel unilaterally or mutually. For the mutual close case, first, it sends a \textit{shutdown} message to the bridge LN node to initiate the closing. If there are no pending HTLCs in the channel, the bridge LN node replies with a \textit{shutdown} message. Now, the LN gateway and the bridge LN node start negotiating on the channel closing fee. Basically, both parties offer each other a fee that they think is fair until they both agree on the same fee. Each offer is done through a \textit{closing\_signed} message which includes the offering party's signature and the offered fee amount. When the LN gateway and the bridge LN node eventually agree on a fee, the resulting closing transaction will be broadcast by the LN gateway. Since the LN gateway only has the bridge LN node's signature for this closing transaction, it needs to also get the IoT device's signature before it can broadcast it. Thus, we propose the LN gateway to send a \textit{ClosingTxRequest} message to the IoT device which has the following fields: \textit{Type: ClosingTxRequest, Closing Transaction}. Upon receiving the request, the IoT device generates a signature for the closing transaction and sends it to the LN gateway in a \textit{ClosingTxSigned} message that has the following fields: \textit{Type: ClosingTxSigned, Signature}. With this signature, the LN gateway can now broadcast the closing transaction to the Bitcoin network and close the channel. The on-chain fee of this transaction is paid by the IoT device by deducting the fee from its balance on the channel. We illustrated the protocol steps for this mutual close case in Fig. \ref{fig:IoTclosechannel}.
    
    \vspace{1mm}
    \textbf{Unilateral Close}: For the unilateral close case, the LN gateway sends its most recent commitment transaction to the IoT device in a \textit{SignTxRequest} message which has the following fields: \textit{Type: SignTxRequest, Commitment Transaction}. Upon receiving the message, the IoT device generates a signature for this commitment transaction and sends it to the LN gateway in a \textit{TxSigned} message that has the following fields: \textit{Type: TxSigned, Signature}. After the LN gateway receives the signature, it can broadcast the commitment transaction to the Bitcoin network to close the channel. Similarly, the on-chain fee is paid by the IoT device.

% \end{itemize}

\subsubsection{LN gateway channel closure}
\label{sec:gatewaychannelclose}

The LN gateway can also close the channel it opened to the bridge LN node due to the IoT device's request. The steps are very similar to the IoT channel closure case with minor differences. We explain it below:

% \begin{itemize}[leftmargin=*]
    \vspace{1mm}
    \textbf{LN Gateway Channel Closing Request}: The LN gateway sends a \textit{ChannelClosingRequest} message to the IoT device.
    
    \vspace{1mm}
    \textbf{Closing the Channel}: The LN gateway can close the channel unilaterally or mutually and it needs the IoT device's signature to be able to close the channel. Therefore, it follows the exact same steps given for the IoT channel closure case above. The only difference is the on-chain fee part. This time, channel closing is requested by the LN gateway; therefore, it pays the on-chain fee. The same mechanism is used: the on-chain fee is deducted from the LN gateway's balance on the channel. There is a chance that the LN gateway might not have enough balance in the channel to cover the on-chain fee. Thus, we propose that the LN gateway does not attempt to close the channel without collecting enough service fees on the channel.
    
% \end{itemize}

\subsubsection{Bridge LN node channel closure}

Similar to the LN gateway channel closure case, the bridge LN node can close the channel unilaterally or mutually. In both scenarios, the LN gateway will learn about closure of the channel. We propose that the LN gateway notifies the IoT device about the channel closure by sending it a \textit{ChannelClosed} message. This serves as a notification to the IoT device so that it does not attempt to use the channel in the future.

\subsection{Changes to LN's BOLT \#3}
\label{sec:changescripts}

We explained in Section \ref{sec:bolt3} the BOLT \#3 specification. Since the channels were made 3-of-3 multisignature with the introduction of the IoT device, it requires changes to the LN's Bitcoin scripts. In this section, we show the proposed changes to the funding transaction output, the commitment transactions, and the HTLC transactions.

\vspace{1mm}
\textbf{Changes to the Funding Transaction Output}:
Instead of sending the funds to a 2-of-2 multisignature address, we propose to send them to a 3-of-3 multisignature address. Thus, the new proposed witness script of the funding transaction output is: \texttt{3 <pubkey1> <pubkey2> <pubkey3> 3 OP\_CHECKMULTISIG}.

\vspace{1mm}
\textbf{Changes to the Commitment Transaction Inputs}:
We propose to change the commitment transaction input witness to: \texttt{0 <sig\_for\_pubkey1> <sig\_for\_pubkey2> <sig\_for\_pubkey3>} as now 3 signatures are needed instead of 2.

\vspace{1mm}
\textbf{Addition of \texttt{to\_IoT} Output to the Commitment Transactions}:
The LN gateway is not funding the channel. Instead, the channel is funded by the IoT device. Therefore, the IoT device needs its own output in commitment transactions. Thus, we propose adding a \texttt{to\_IoT} output to the LN gateway's and bridge LN node's commitment transactions. This output pays to the \texttt{IoT\_pubkey} that the IoT device can spend with the witness \texttt{<IoT\_sig>}.

\vspace{1mm}
\textbf{Changes to \texttt{to\_local} Output of the Commitment Transactions}:
Since the channel is funded by the IoT device, the \texttt{to\_local} output in the LN gateway's commitment transaction only holds the LN gateway's service fees which are sent to the LN gateway's \texttt{<local\_delayedpubkey>}. The bridge LN node's \texttt{to\_local} output is not modified.

\vspace{1mm}
\textbf{Changes to \texttt{to\_remote} Output of the Commitment Transactions}:
There are no changes to the \texttt{to\_remote} outputs. The LN gateway's \texttt{to\_remote} output pays to the bridge LN node's \texttt{remotepubkey}, the bridge LN node's \texttt{to\_remote} output pays to the LN gateway's \texttt{remotepubkey}.

\vspace{1mm}
\textbf{Changes to Offered HTLC Outputs of the Commitment Transactions}:
The witness script of this output normally has \texttt{OP\_DROP 2 OP\_SWAP <local\_htlcpubkey> 2 OP\_CHECKMULTISIG} which sends the funds to the local node with the HTLC-timeout transaction. For the LN gateway's offered HTLC outputs, the local node is the IoT device instead of the LN gateway itself, thus \texttt{local\_htlcpubkey} is changed to \texttt{IoT\_htlcpubkey}. Our protocol does not support offered HTLC outputs for the bridge LN node's commitment transaction. This is because our protocol only supports payments from the IoT device to destination LN nodes. This is a limitation which we plan to address in the future.

\vspace{1mm}
\textbf{Changes to Received HTLC Outputs of the Commitment Transactions}:
The current design does not support received HTLC outputs for the LN gateway's commitment transaction as the IoT device cannot receive payments on the channel. On the other hand, the bridge LN node's commitment transaction supports received HTLC outputs and we do not propose any changes.

\vspace{1mm}
\textbf{Changes to HTLC-Timeout and HTLC-Success Transactions}:
As explained above, we do not support HTLC-success transactions for the LN gateway's commitment transaction as the IoT device cannot receive payment on the channel. Similarly, HTLC-timeout transactions are not supported for the bridge LN node's commitment transaction. For the HTLC-timeout transactions in the LN gateway's commitment transaction, we propose having 3 signatures instead of 2. Thus, the new transaction input witness will be: \texttt{0 <remotehtlcsig> <localhtlcsig> <IoThtlcsig> <>}. For the HTLC-success transactions in the bridge LN node's commitment transaction, we again propose having 3 signatures instead of 2. The new transaction input witness will be: \texttt{0 <remotehtlcsig> <IoThtlcsig> <localhtlcsig> <payment\_preimage>}. Additionally, the \texttt{local\_delayedpubkey} in the witness script for the output of the HTLC-timeout transaction in the LN gateway's commitment transaction is changed to \texttt{IoT\_delayedpubkey}.

\subsection{Handling revoked state broadcasts}
We briefly mentioned in Section \ref{sec:LNmechanisms} that Alice and Bob can attempt to cheat by broadcasting revoked channel states to the blockchain. LN uses \textit{timelocks} to address this issue. The idea is not to let the broadcasting party spend its funds immediately while letting the counterparty do so. Since we proposed changes to the LN protocol, it requires revisiting the revoked state cases. Specifically, the IoT device does not store any commitment transactions nor revocation keys. It is only involved in signing operations. Therefore, this situation should not result in any loss of funds for the IoT device in possible cheating attempts by other channel parties. We examine the possible revoked state broadcast cases by the LN gateway and bridge LN node separately below and show how both cases are handled properly.

\subsubsection{Revoked state broadcast by the LN gateway}
The LN gateway can broadcast revoked commitment transactions to the blockchain since they are already signed by everyone. This attempt has 2 possible outcomes: 1) The bridge LN node was offline for long enough to not realize the LN gateway was cheating. Thus, it loses some or all of its funds in the channel depending on the broadcast old state. 2) The bridge LN node was online during the LN gateway's cheating attempt thus, sweeps all the funds in the channel using the revocation private key of the respective old state.

The first scenario is the famous \textit{being offline} issue for the LN nodes. All existing LN nodes are vulnerable to this attack when they are offline for extended periods of time \cite{offlinelnnodes}. Therefore, it is not specific to our protocol and addressing it is beyond the scope of this paper\footnote{Watchtowers \cite{watchtowers} were proposed to protect LN nodes against this threat.}.

\begin{figure}[h]
    \centering
    % \vspace{-3mm}
    \includegraphics[width=\linewidth]{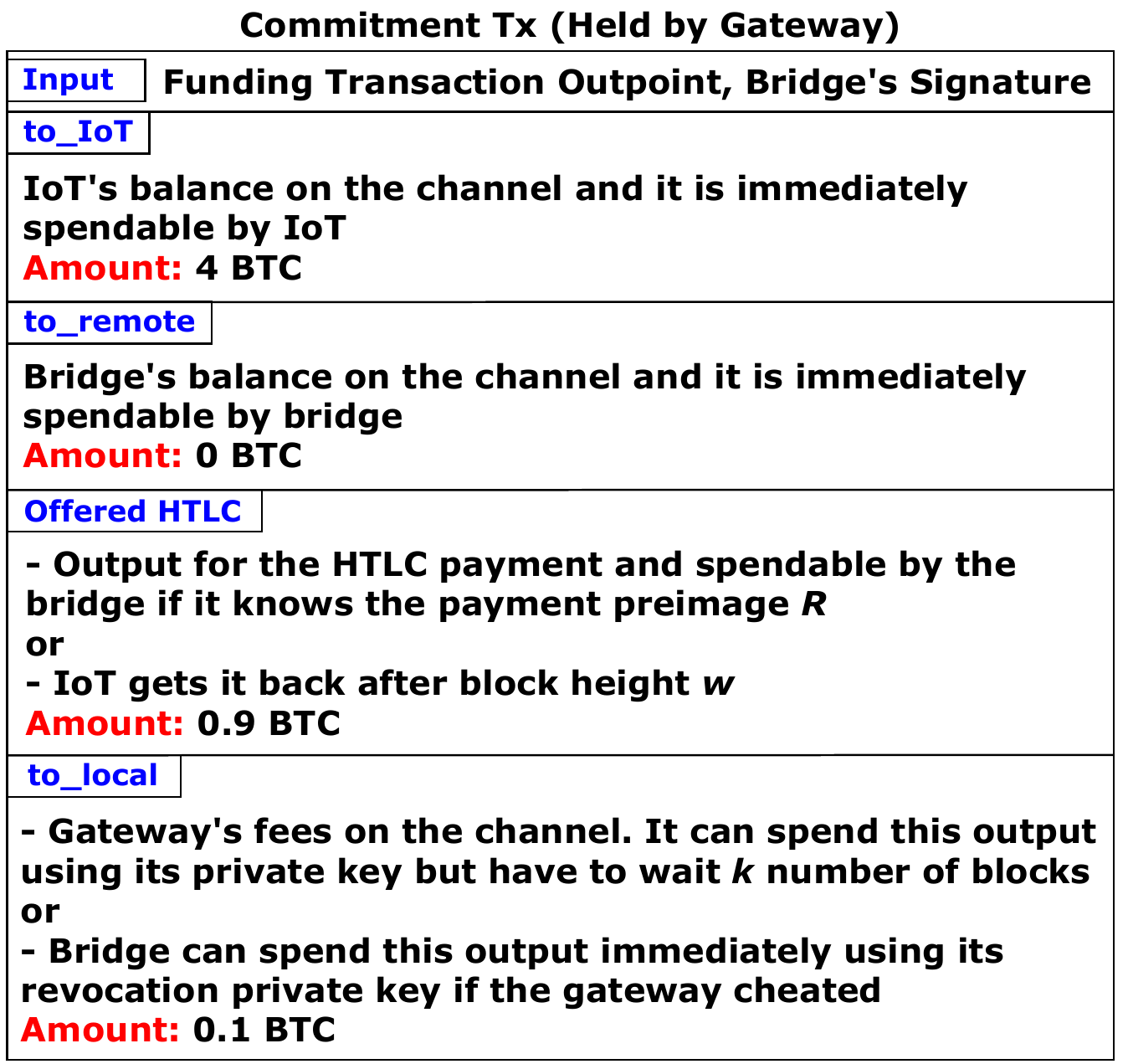}
    \vspace{-5mm}
    \caption{An illustration of the commitment transaction stored at the LN gateway with the IoT output and the fee output modified (\texttt{to\_IoT} and \texttt{to\_local}, respectively). This commitment transaction is generated after the following events: A channel with 5 BTC capacity is opened; the IoT device initiated a 1 BTC payment to a destination; the LN gateway charged the IoT device a service fee of 0.1 BTC.}
    % \vspace{-2mm}
    \label{fig:gatewaycommitment}
\end{figure}

However, the second scenario jeopardizes the IoT device's funds if not addressed. To handle this case, we propose two modifications to the LN gateway's commitment transaction. The first modification is for the \texttt{to\_IoT} output at which the IoT device's funds are held. We propose that even after a failed cheating attempt by the LN gateway, the bridge LN node cannot spend the \texttt{to\_IoT} output. In other words, we propose to make this output spendable only by the IoT device at all times. In this way, the IoT device's funds in the channel will be protected. The second modification is proposed for the fee output (\texttt{to\_local}) of the LN gateway. Normally, this output is spendable by the LN gateway only. We propose to turn it into a conditional output such that, if the LN gateway gets caught while cheating, the bridge LN node can spend this output using the revocation private key. In other words, the LN gateway will lose the fees it collected on the channel if it gets caught while cheating. This modification disincentivizes the LN gateway from attempting to cheat therefore addresses the revoked state broadcast issue. Consequently, with these 2 modifications, the IoT device's funds are protected and the LN gateway is disincentivized from broadcasting revoked states. We illustrate the LN gateway's modified commitment transaction in Fig. \ref{fig:gatewaycommitment}.

\subsubsection{Revoked state broadcast by the bridge LN node}
Similar to the LN gateway, the bridge LN node can also broadcast its revoked commitment transactions to the blockchain in an attempt to cheat. Depending on the published old state, it might benefit the bridge LN node. However, this attempt will be successful only if the LN gateway is offline during the attempt. This brings us back to the \textit{being offline} issue. Basically, it is the LN gateway's responsibility to stay online and protect itself against this attack. As this is a general LN issue, it is beyond the scope of this paper.

\section{Security Analysis}
\label{sec:threatanalysis}
In this section, we present how the attacks mentioned in Section \ref{sec:threatmodel} are mitigated in our approach.

\vspace{1mm}
\noindent \textbf{Threat 1: Revoked State Broadcasts}: For the attack where the LN gateway broadcasts a revoked state, our approach proposed punishing the LN gateway. Basically, the LN gateway loses the service fees it collected on the channel to bridge the LN node. With our punishment addition, the LN gateway is disincentivized from broadcasting a revoked state. Additionally, we proposed a protection mechanism for the IoT device's funds on the channel. In this way, even when the bridge LN node catches the LN gateway while cheating, the IoT device does not lose any funds. The other cases where the parties might lose funds because of being offline are not handled as they are general LN issues that are not related to our protocol.

\vspace{1mm}
\noindent \textbf{Threat 2: Spending IoT Device's Funds:}
Using 3-of-3 multisignature channels secure the IoT device's funds in the channel since the LN gateway cannot spend them without getting the IoT device's cryptographic signature first. As shown in Fig. \ref{fig:sendpayment}, the LN gateway sends the newly generated bridge LN node's version of the commitment transaction to the IoT device for signing in step 4. Without performing this step, the LN gateway cannot get a signature from the bridge LN node for its own version of the commitment transaction. Therefore, it is apparent that the LN gateway will not be able to complete a successful payment without getting a signature from the IoT device. On the other hand, if we were to use LN's original 2-of-2 multisignature channels in our system, the LN gateway could move the funds without needing a signature from the IoT device. Because, with a 2-of-2 multisignature channel, the LN gateway could send its own signature to the bridge LN node which would be enough for the bridge LN node to be able to spend its commitment transaction. Since the LN gateway cannot spend the IoT device's funds in the channel at its own will, the IoT device's funds are always protected and can be only spent when the IoT device provides its signature. Consequently, the usage of 3-of-3 multisignature channels protects the IoT device's funds from getting unwillingly spent by the LN gateway.

\section{Evaluation}
\label{sec:evaluation}

In this section, we explain our experiment setup and present the performance results.

\subsection{Experiment setup and metrics}

To evaluate the proposed protocol, we created a setup where an IoT device connects to an LN gateway to send payments on the LN. To mimic the IoT device, we used a Raspberry Pi 3 Model B v1.2 and the LN gateway was set up on a desktop computer with an Intel(R) Xeon(R) CPU E5-2630 v4 and 32 GB of RAM. This desktop computer was in a remote location different from that of Pi's. For the full Bitcoin node installation, we used \textit{bitcoind} \cite{bitcoind} which is one of the implementations of the Bitcoin protocol. For the LN node, we used \textit{lnd v0.11.0-beta.rc1} from Lightning Labs \cite{lnd} which is a complete implementation of the LN protocol. Python was used to implement the protocol.

We used IEEE 802.11n (WiFi) and Bluetooth Low Energy (BLE 4.0) to exchange protocol messages between the Raspberry Pi and the LN gateway. In the WiFi scenario, we created a server \& client TCP socket application in Python. With this, Raspberry Pi and the remote desktop computer communicated with each other. Raspberry Pi was connected to the Internet through a regular Internet modem which acted as the IoT gateway. For Bluetooth experiments, we first paired the Raspberry Pi and the Bluetooth adapter of the IoT gateway. We used a laptop computer as the IoT gateway which had a Bluetooth adapter. Using Python's \textit{bluetooth} library, Raspberry Pi and the laptop computer communicated with each other. The laptop computer was programmed to talk to the LN gateway over the same TCP socket application that was set up. In both cases, the LN gateway used gRPC API \cite{grpc} of \textit{lnd} to communicate with the LN node that was running on it. 

To assess the performance of our protocol, we used the following metrics: 1) \textit{Time} which refers to the total computational and communication delays of the proposed protocol; and 2)  \textit{Cost} which refers to the total monetary cost associated with sending payments using the LN gateway.

To compare our approach to a baseline, we considered the case where the LN gateway sends the payments to a destination by itself. In other words, no IoT device is present, and all LN tasks are solely performed by the LN gateway.

\subsection{Computational and communication delays}
\label{sec:overheadresults}

We first assessed the computational and communication delays of our proposed protocol. The computational delay of running the protocol on a Raspberry Pi comes from the AES encryption of the protocol messages and HMAC calculations. We used Python's \textit{pycrypto} library to encrypt the protocol messages with AES-256 encryption. The encrypted data size for the messages was 24 bytes. For the HMAC calculations, we used \textit{hmac module} in Python. The delays for these operations are shown in Table \ref{tab:comp}. As can be seen, they are negligible.

\begin{table}[h]
%  \vspace{-5mm}
  \begin{center}
%   \vspace{-3mm}
    \caption{Computational delays on the IoT device}
     \vspace{-2mm}
    \label{tab:comp}
    \resizebox{0.8\linewidth}{!}{
    \begin{tabular}{|c|c|c|}
    \hline
       \textbf{AES Encryption} &  \textbf{HMAC Calculation} & \textbf{Total} \\
     \hline
        15 ms             &       $<$ 1 ms    & 15 ms   \\
     \hline
    \end{tabular}
    }
   \vspace{-3mm}
  \end{center}
\end{table}

We then measured the communication delays which are used in other experiments to evaluate the timeliness of the protocol. We define the communication delay as the delay of sending a protocol message from the Raspberry Pi to the IoT gateway and receiving an acknowledgment for that message from the IoT gateway or vice versa. This delay is also called the round-trip time of a protocol message. When WiFi was used for the connection, the measured round-trip delay of a message was approximately \textit{9 ms}. When Bluetooth was used, the same delay was \textit{0.8 seconds}. As expected, the message exchanges with Bluetooth are much slower compared to WiFi which is related to the bandwidth difference between the two technologies.

\subsection{Toll payment use case evaluation of the protocol}

In this part, we consider an example case where real-time response is critical. We assume a toll application where cars pass through a toll gate and pay the toll without stopping. For this, wireless technologies are used. The cars that enter the communication range of the toll gate's wireless system immediately initiates a payment to the toll company's LN node through the toll's LN gateway which is running on the cloud. Cars are notified upon a successful payment. In order for this process to work, payment sending has to be completed while the car is still in the communication range of the toll gate's wireless system.

As can be seen in Fig. \ref{fig:sendpayment}, there are 4 protocol message exchanges between the IoT device and the LN gateway in our payment sending protocol. For this toll example, in each protocol message exchange, there are 2 corresponding communication delays which are between the car and the IoT gateway and between the IoT gateway and the LN gateway running on the cloud. We know the communication delays between the car and the IoT gateway from the previous section which were 9 ms and 0.8 seconds for WiFi and Bluetooth respectively. The delay of the communication between the IoT gateway and the LN gateway running on the cloud on the other hand will be the delay of a regular TCP communication between two computers. We measured an average of 123 ms delay for this communication. The actual LN payment on the other hand was sent in 2.1 seconds which is an average value calculated from 30 separate payments sent at different times throughout the day. There is also the 15 ms computational delay at the car for each protocol message it generates. Eventually, the \textit{total payment sending time for WiFi was 2.658 seconds while BLE had 5.822 seconds}. 

We mentioned earlier that 802.11n and BLE were used for the measurements. The advertised range of 802.11n is approximately 250 meters \cite{matsumoto2009performance} and the advertised range of BLE is around 220 meters \cite{BLERange}. If cars pass through the toll gate with a speed of 50 miles per hour, there is around 11 seconds with WiFi and 10 seconds with Bluetooth available for them to complete the protocol message exchanges with the LN gateway for a successful toll payment. The results for varying vehicle speeds are shown in Table \ref{tab:toll}. As can be seen, for both WiFi and Bluetooth cases, our protocol meets the deadlines even under a high speed of 80 mph. Even in the case of Bluetooth where the data rates are low, the deadline can be met due to the long communication ranges of Bluetooth 4.0. If different technologies with limited ranges are used, cars' speed should be enforced accordingly.

\begin{table}[h]
%  \vspace{-3mm}
  \begin{center}
    \caption{Available time under different speeds to make a successful toll payment with WiFi and Bluetooth}
     \vspace{-2mm}
    \label{tab:toll}
    \resizebox{\linewidth}{!}{
    \begin{tabular}{|c|c|c|c|c|}
      \hline
      &   \multicolumn{2}{c|}{\textbf{WiFi}}   &  \multicolumn{2}{c|}{\textbf{Bluetooth}}   \\ 
     \hline 
     \textbf{Vehicle Speed}  &  \textbf{Available Time} & \textbf{Satisfied?} & \textbf{Available Time} & \textbf{Satisfied?} \\
     \hline
       50 mph  &  11.2 s   &  Yes  &  9.8 s  & Yes \\ 
      \hline     
       60 mph  &  9.3 s   &  Yes  &  8.2 s  & Yes \\ 
      \hline
       80 mph  &  7 s     &  Yes  &  6.2 s  & Yes \\
     \hline
     
    \end{tabular}
  }
  \vspace{-6mm}
  \end{center}
\end{table}

\subsection{Coffee shop example}
As another real-life example, let us consider a case where customers pay for coffee at a coffee shop with their smartwatches using our protocol. This payment is less time-critical compared to the toll payments, since the customers can wait by the cashier until they get the payment confirmation from the coffee shop's LN node. We already measured the total payment sending time for WiFi and Bluetooth cases which were 2.658 seconds and 5.822 seconds, respectively. Therefore, again in this example, the use of our protocol through the WiFi and Bluetooth wireless technologies is feasible since the customers can wait for more than 5.822 seconds for payment confirmations.

As an alternative, customers can also pay for the coffee using an existing device in the coffee shop that is connected to the LN and ready to send payments using its existing LN channels. In this case, however, there is no IoT device involved; therefore, no wireless communication delays. We call this option \textit{No IoT Case}. Since there are no communication delays in this scenario, payment sending only takes 2.1 seconds. When compared to the case where the coffee is paid with a smartwatch, the difference in the payment sending time is the communication delays for the protocol message exchanges. The results of this coffee shop example are summarized in Table \ref{tab:coffee}.

\begin{table}[h]
 %   \vspace{-3mm}
    \caption{Total payment sending time comparison of all three cases for the coffee shop example}
    \vspace{-2mm}
    \label{tab:coffee}
    \resizebox{\linewidth}{!}{
    \begin{tabular}{|c|c|c|}
      \hline
      \textbf{Our Approach - WiFi}  & \textbf{Our Approach - BLE} & \textbf{No IoT Case}  \\ \hline
              2.66 seconds & 5.82 seconds & 2.1 seconds   \\ \hline
    \end{tabular}
    }
    % \vspace{-6mm}
\end{table}

\subsection{Cost analysis}
We now investigate the total monthly payment sending cost of the IoT device. The only associated cost of the payment sending comes from the fees the LN gateway charges when it sends payments for the IoT device. We assume that the fee that will be charged totally depends on the LN gateway and specific use case of the service (i.e., paying for toll, paying for coffee, etc.).
For the toll example, let us assume that a car passes through the toll 2 times a day. If the toll charges \$1.5 per pass, the car pays \$3 a day. The LN gateway also charges a \textit{\%k} fee on top of the toll. If we take \textit{k}=10, then the car pays \$3.3 in total, \$0.3 of which goes to the LN gateway. The LN gateway's fee includes the LN's payment routing fees which are usually around a few satoshi per payment \cite{lnroutingfees}. Then in one month, the car will pay \$9 to the LN gateway for the fees. While it depends on the driver, we believe that the reflected cost is negligible considering the comfort of the fast toll payments.

\section{Conclusion}
\label{sec:conclusion}

In this paper, we proposed a secure and efficient protocol for enabling IoT devices to use Bitcoin's LN for sending payments. By modifying LN's existing peer protocol and on-chain Bitcoin transactions, a third peer (i.e. IoT device) was added to the LN channels. The purpose was to enable resource-constrained IoT devices that normally cannot interact with LN to interact with it and perform micro-payments with other users. The IoT device's interactions with LN are achieved through a gateway node that has access to LN and thus can provide LN services to it in return for a fee. In order to prevent possible threats that might arise from broadcasting old states, LN's commitment transactions were modified. Our evaluation results showed that the proposed protocol enables LN payments for the IoT devices with negligible delays.

\bibliographystyle{unsrt}
\bibliography{references}

\begin{thebibliography}{10}

\bibitem{li2015internet}
Shancang Li, Li~Da~Xu, and Shanshan Zhao.
\newblock The internet of things: a survey.
\newblock {\em Information Systems Frontiers}, 17(2):243--259, 2015.

\bibitem{pavsalic2016vehicle}
Dra{\v{z}}en Pa{\v{s}}ali{\'c}, Branimir Cviji{\'c}, Du{\v{s}}anka Bundalo,
  Zlatko Bundalo, and Radovan Stojanovi{\'c}.
\newblock Vehicle toll payment system based on internet of things concept.
\newblock In {\em 2016 5th Mediterranean Conference on Embedded Computing
  (MECO)}, pages 485--488. IEEE, 2016.

\bibitem{mercan2021cryptocurrency}
Suat Mercan, Ahmet Kurt, Enes Erdin, and Kemal Akkaya.
\newblock Cryptocurrency solutions to enable micro-payments in consumer {IoT}.
\newblock {\em IEEE Consumer Electronics Magazine}, 2021.

\bibitem{kurt2020lnbot}
Ahmet Kurt, Enes Erdin, Mumin Cebe, Kemal Akkaya, and A~Selcuk Uluagac.
\newblock {LNBot}: A covert hybrid botnet on bitcoin lightning network for fun
  and profit.
\newblock In {\em European Symposium on Research in Computer Security}, pages
  734--755. Springer, 2020.

\bibitem{nakamoto2019bitcoin}
Satoshi Nakamoto.
\newblock Bitcoin: A peer-to-peer electronic cash system.
\newblock Technical report, 2008.
\newblock \url{https://bitcoin.org/bitcoin.pdf}.

\bibitem{wood2014ethereum}
Gavin Wood.
\newblock Ethereum: A secure decentralised generalised transaction ledger.
\newblock Technical report, 2014.
\newblock \url{https://github.com/ethereum/yellowpaper}.

\bibitem{zhou2020solutions}
Qiheng Zhou, Huawei Huang, Zibin Zheng, and Jing Bian.
\newblock Solutions to scalability of blockchain: A survey.
\newblock {\em IEEE Access}, 8:16440--16455, 2020.

\bibitem{decker2015fast}
Christian Decker and Roger Wattenhofer.
\newblock A fast and scalable payment network with bitcoin duplex micropayment
  channels.
\newblock In {\em Symposium on Self-Stabilizing Systems}, pages 3--18.
  Springer, 2015.

\bibitem{poon2016bitcoin}
Joseph Poon and Thaddeus Dryja.
\newblock The bitcoin lightning network: Scalable off-chain instant payments.
\newblock Technical report, 2016.
\newblock \url{http://lightning.network/lightning-network-paper.pdf}.

\bibitem{lniotproblem}
Kenichi Kurimoto.
\newblock Lightning network x {IoT(LoT)}; potential, challenges and solutions,
  accessed 2021-03.
\newblock
  \url{https://medium.com/nayuta-en/lightning-network-x-iot-lot-potential-challenges-and-solutions-6e4d8b4c252a}.

\bibitem{hannon2019bitcoin}
Christopher Hannon and Dong Jin.
\newblock Bitcoin payment-channels for resource limited {IoT} devices.
\newblock In {\em Proceedings of the International Conference on Omni-Layer
  Intelligent Systems}, pages 50--57, 2019.

\bibitem{robert2020enhanced}
J{\'e}r{\'e}my Robert, Sylvain Kubler, and Sankalp Ghatpande.
\newblock Enhanced lightning network (off-chain)-based micropayment in {IoT}
  ecosystems.
\newblock {\em Future Generation Computer Systems}, 2020.

\bibitem{pouraghily2019lightweight}
Arman Pouraghily and Tilman Wolf.
\newblock A lightweight payment verification protocol for blockchain
  transactions on {IoT} devices.
\newblock In {\em 2019 International Conference on Computing, Networking and
  Communications (ICNC)}, pages 617--623. IEEE, 2019.

\bibitem{raiden}
Brainbot Labs.
\newblock $\mu$raiden - a payment channel framework for fast \& free off-chain
  {ERC20} token transfers, accessed 2021-03.
\newblock \url{https://raiden.network/micro.html}.

\bibitem{profentzas2020tinyevm}
Christos Profentzas, Magnus Almgren, and Olaf Landsiedel.
\newblock {TinyEVM}: Off-chain smart contracts on low-power {IoT} devices.
\newblock In {\em 2020 IEEE 40th International Conference on Distributed
  Computing Systems (ICDCS)}, pages 507--518. IEEE, 2020.

\bibitem{li2020data}
Dunfeng Li, Yong Feng, Yao Xiao, Mingjing Tang, and Xiaodong Fu.
\newblock A data trading scheme based on payment channel network for internet
  of things.
\newblock In {\em International Conference on Blockchain and Trustworthy
  Systems}, pages 319--332. Springer, 2020.

\bibitem{tapas2020p4uiot}
Nachiket Tapas, Yechiav Yitzchak, Francesco Longo, Antonio Puliafito, and Asaf
  Shabtai.
\newblock {P4UIoT}: Pay-per-piece patch update delivery for {IoT} using gradual
  release.
\newblock {\em Sensors}, 20(7):2156, 2020.

\bibitem{neutrino}
Lightning Labs.
\newblock Neutrino: Privacy-preserving bitcoin light client, accessed 2021-03.
\newblock \url{https://github.com/lightninglabs/neutrino}.

\bibitem{breez}
Breez.
\newblock Breez mobile client, accessed 2021-03.
\newblock \url{https://github.com/breez/breezmobile}.

\bibitem{lnd}
Lightning Labs.
\newblock Lightning network daemon, accessed 2021-03.
\newblock \url{https://github.com/lightningnetwork/lnd}.

\bibitem{kurt2021enabling}
Ahmet Kurt, Suat Mercan, Enes Erdin, and Kemal Akkaya.
\newblock Enabling micro-payments on {IoT} devices using bitcoin lightning
  network.
\newblock In {\em 2021 IEEE International Conference on Blockchain and
  Cryptocurrency (ICBC)}, pages 1--3. IEEE, 2021.

\bibitem{lnlaunch}
Lightning Labs.
\newblock Announcing our first lightning mainnet release, lnd 0.4-beta!,
  accessed 2021-03.
\newblock
  \url{https://blog.lightning.engineering/announcement/2018/03/15/lnd-beta.html}.

\bibitem{smartcontract}
Bitcoin Wiki.
\newblock Contract, accessed 2021-03.
\newblock \url{https://en.bitcoin.it/wiki/Contract}.

\bibitem{1ml}
1ml.com.
\newblock Lightning network search and analysis engine, accessed 2021-03.
\newblock \url{https://1ml.com/}.

\bibitem{lntopology}
Bryan Vu.
\newblock Exploring lightning network routing, accessed 2021-03.
\newblock
  \url{https://blog.lightning.engineering/posts/2018/05/30/routing.html}.

\bibitem{bolt2}
Lightning Labs.
\newblock {BOLT} \#2: Peer protocol for channel management, accessed 2021-03.
\newblock
  \url{https://github.com/lightningnetwork/lightning-rfc/blob/master/02-peer-protocol.md}.

\bibitem{bolt3}
Lightning Labs.
\newblock {BOLT} \#3: Bitcoin transaction and script formats, accessed 2021-03.
\newblock
  \url{https://github.com/lightningnetwork/lightning-rfc/blob/master/03-transactions.md}.

\bibitem{offlinelnnodes}
Ichiro Kuwahara.
\newblock Operating lightning \#0001 — basic challenges in operating lighting
  network nodes, accessed 2021-03.
\newblock
  \url{https://medium.com/crypto-garage/operating-lightning-0001-basic-challenges-in-operating-lighting-network-nodes-d04386ac931a}.

\bibitem{watchtowers}
lnd.
\newblock Private altruist watchtowers, accessed 2021-03.
\newblock
  \url{https://github.com/lightningnetwork/lnd/blob/master/docs/watchtower.md}.

\bibitem{bitcoind}
bitcoin.org.
\newblock Running a full node, accessed 2021-03.
\newblock \url{https://bitcoin.org/en/full-node}.

\bibitem{grpc}
Lightning Labs.
\newblock {LND gRPC API} reference, accessed 2021-03.
\newblock \url{https://api.lightning.community/}.

\bibitem{matsumoto2009performance}
Akira Matsumoto, Kouichi Yoshimura, Stefan Aust, Tetsuya Ito, and Yoshihisa
  Kondo.
\newblock Performance evaluation of {IEEE} 802.11n devices for vehicular
  networks.
\newblock In {\em 2009 IEEE 34th Conference on Local Computer Networks}, pages
  669--670. IEEE, 2009.

\bibitem{BLERange}
Heikki Karvonen, Carlos Pomalaza-R{\'a}ez, Konstantin Mikhaylov, Matti
  H{\"a}m{\"a}l{\"a}inen, and Jari Iinatti.
\newblock Experimental performance evaluation of {BLE} 4 versus {BLE} 5 in
  indoors and outdoors scenarios.
\newblock In {\em Advances in Body Area Networks I}, pages 235--251, 2019.

\bibitem{lnroutingfees}
BitMEX Research.
\newblock The lightning network (part 2) – routing fee economics, accessed
  2021-03.
\newblock
  \url{https://blog.bitmex.com/the-lightning-network-part-2-routing-fee-economics/}.

\end{thebibliography}

\newpage
\section*{Authors}

\label{sec:auth}

\begin{wrapfigure}{l}{.13\textwidth}\includegraphics[width=0.35\columnwidth]{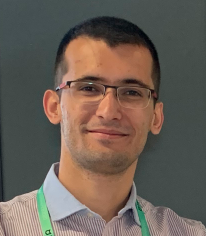}\end{wrapfigure}
\textbf{Ahmet Kurt} received two B.S. degrees from Antalya Bilim University, Antalya, Turkey in 2018. He is currently pursuing a Ph.D. degree in Electrical and Computer Engineering with the Florida International University, Miami, United States. His current research interests include Bitcoin's lightning network, Bitcoin and payment channel networks.

\vspace{1em}
\ITUpar

\begin{wrapfigure}{l}{.13\textwidth}\includegraphics[width=0.35\columnwidth]{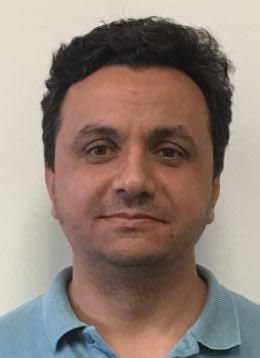}\end{wrapfigure}
\textbf{Suat Mercan} is a postdoctoral researcher at Florida International University. He received his Ph.D. degree in Computer Science from the University of Nevada, Reno in 2011 and his M.S degree in Electrical and Computer Engineering from the University of South Alabama in 2007. His main research interests are blockchain, payment channel and peer-to-peer networks, cybersecurity, digital forensics, and content delivery.

\vspace{1em}
\ITUpar

\begin{wrapfigure}{l}{.13\textwidth}\includegraphics[width=0.35\columnwidth]{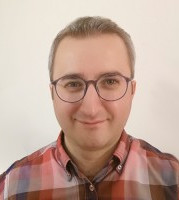}\end{wrapfigure}
\textbf{Enes Erdin} is an Assistant Professor at University of Central Arkansas. He received his Ph.D. degree in Electrical and Computer Engineering from Florida International University and he was an NSF CyberCorps Fellow. He conducts research in the areas of hardware security, blockchain technology, and cyber-physical systems.

\vspace{1em}
\ITUpar

\begin{wrapfigure}{l}{.13\textwidth}\includegraphics[width=0.35\columnwidth]{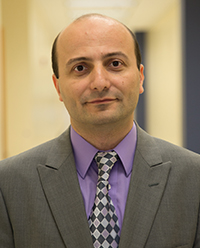}\end{wrapfigure}
\textbf{Kemal Akkaya} is a professor in the Department of Electrical and Computer Engineering at Florida International University. He leads the Advanced Wireless and Security Lab and is an area editor of the Elsevier Ad Hoc Networks Journal. His current research interests include security and privacy, and protocol design. He has published over 120 papers in peer-reviewed journals and conferences. He received the ``Top Cited'' article award from Elsevier in 2010.

\end{document}